\documentclass[twocolumn,prA,showpacs,amsmath,amssymb]{revtex4}
\usepackage{graphicx}
\usepackage{amsfonts}

\def\be{\begin{equation}}
\def\ee{\end{equation}}
\def\bea{\begin{eqnarray}}
\def\eea{\end{eqnarray}}
\def\bma{\begin{mathletters}}
\def\ema{\end{mathletters}}

\def\0{\overline{0}}

\def\q0{\underline{0}}

\def\C{{\cal C}}

\def\E{{\cal E}}

\def\R{{\cal R}}

\def\tr{\mbox{tr}}
\def\one{\leavevmode\hbox{\small1\normalsize\kern-.33em1}}

\def\bra#1{\langle#1|} \def\ket#1{|#1\rangle}
\def\braket#1#2{\langle#1|#2\rangle}

\begin{document}

\title{Gaussian Operations and Privacy}
\author{Miguel Navascu\'es and Antonio Ac\'{\i}n}

\affiliation{ICFO-Institut de Ci\`encies Fot\`oniques, Jordi
Girona 29, Edifici Nexus II, E-08034 Barcelona, Spain}
\date{\today}


\begin{abstract}
We consider the possibilities offered by Gaussian states and
operations for two honest parties, Alice and Bob, to obtain
privacy against a third eavesdropping party, Eve. We first extend
the security analysis of the protocol proposed in M. Navascu\'es
{\sl et al.}, Phys. Rev. Lett. {\bf 94}, 010502 (2005). Then, we
prove that a generalized version of this protocol does not allow
to distill a secret key out of bound entangled Gaussian states.
\end{abstract}


\maketitle

\section{Introduction}
\label{intro}

The study of those tasks that can be achieved by processing
information encoded on quantum states is the main scope of Quantum
Information Theory (QIT). The basic unit for quantum information
is the so-called quantum bit, or {\em qubit}, namely a
two-dimensional quantum system. Moreover, quantum correlations, or
entanglement, constitute a key resource in QIT, their unit being
the entangled bit or {\em ebit}. In general, any (quantum)
information task can be seen as an inter-conversion of resources.
For instance, quantum teleportation \cite{telep} can be summarized
as the process transforming 1 ebit + 2 bits $\rightarrow$ 1 qubit,
while dense coding \cite{denscod} corresponds to the
transformation 1 ebit + 1 qubit $\rightarrow$ 2 bits. Moving to
cryptography, secret bits are a fundamental resource. These are
perfectly correlated and random bits shared by two honest parties,
Alice and Bob, about which a third dishonest party, Eve, has no
information. Any quantum key distribution protocol can be seen as
the process of distributing secret bits through an insecure
channel by means of quantum states. Therefore, relevant questions
in this context are to identify those quantum states containing
secret correlations and show how to distill these correlations
into a perfect secret key. Indeed, it has recently been shown that
a quantum state contains secret correlations if and only if it is
entangled \cite{AG}.

In these last years, Quantum Information Theory for Continuous
Variables systems has proved to be a very fruitful area, as it
allows theory to connect easily with experiments (for a review,
see \cite{review}). In this case, the information encoding is done
on continuous quantum variables, such as the quadratures of the
electromagnetic field. Recent works have been developed in the aim
of reproducing well-known Quantum Information protocols for
finite-dimensional systems in this new setup. Examples of these
are quantum cryptography \cite{crypt} or quantum teleportation
\cite{teleport}. Interestingly, most of these protocols work using
only Gaussian operations, i.e., operations that transform Gaussian
states into Gaussian states. This is important because Gaussian
operations are easy to implement experimentally with high accuracy
level. A beam splitter or a squeezer are examples of Gaussian
operations, while photon counting constitutes a non-Gaussian
operation. Up to now, non-Gaussian operations are challenging from
an experimental
point of view (see however \cite{Grangier}). 


A significant effort has been devoted to study the possibilities
and limitations Gaussian Operations provide to Quantum Information
protocols. We know, for example, that entanglement distillation of
Gaussian states with Gaussian operations is impossible
\cite{Cirac,Eisert,Fiurasek}. More precisely: although there exist
entangled Gaussian states that are distillable to singlets, the
distillation process requires a non-Gaussian operation. Or, in
other words, the process of converting Gaussian quantum states
into perfect ebits by means of Gaussian local operations and
classical communication (GLOCC) is impossible. However, ebits is
not the only information resource two collaborating parties may
want to establish through quantum states. Actually, distillation
of perfect secret bits by GLOCC is known to be possible from some
Gaussian states \cite{crypt}. Thus, the set of Gaussian states and
operations can sometimes be sufficient for cryptographic
applications. At first sight, this result may seem surprising
taking into account that Gaussian states have a positive Wigner
function, i.e., there is a local variable model that reproduces
the correlations given by Gaussian measurements.


In this article, we analyze the process of extracting secret bits
from several copies of a given Gaussian state when the honest
parties are allowed to perform local Gaussian operations and
communicate classically. In the derivation of all the results, it
is assumed that Alice and Bob share $N$ independent copies of a
known Gaussian state. That is, we do not consider the important
problem of the distribution and estimation of these states. They
simply constitute an initially given resource that the honest
parties will to convert into secret bits. We start reviewing the
results of \cite{mayo}, where it was shown that, provided Eve is
restricted to individual attacks, a secret key can be extracted
from any entanglement distillable state. We extend the security
analysis of this protocol for the case of collective attacks,
giving a necessary and sufficient condition for secret key
distillation. We also show that there is no way in which the
honest parties can attain privacy with our protocol if the initial
state is bound entangled. This is true even if Eve is assumed to
measure her state before any reconciliation process has taken
place. This suggests that Gaussian operations may be useless to
extract a secret key out of bound entangled Gaussian states, in
opposition to the astounding results in \cite{Horo} for
finite-dimensional systems.


The article is organized as follows: Section \ref{revsec} is a
brief introduction to the Gaussian states and Gaussian operations
formalism. The reader familiar with both topics can skip this
part. In Section \ref{cprivacy}, we analyze the limits of the
protocol introduced in \cite{mayo}. In particular, we show that it
allows to prove the security of sufficiently entangled states,
while it fails for any bound entangled state. Section \ref{concl}
is devoted to our conclusions.


\section{Gaussian states and operations}
\label{revsec}

In this article we consider quantum systems of $n$ canonical
degrees of freedom, called modes, belonging to $B({\cal
H}({\mathbb R}^n))$. These are characterized by operators
$(X_1,P_1, ..., X_n, P_n)=(R_1, ..., R_{2n})$ satisfying the
commutation relations $[R_j,R_j]=i(\sigma_n)_{jk}$, where
\begin{equation}
\sigma_n=\oplus_{i=1}^n
\left(\begin{array}{cc}0&1\\-1&0\end{array}\right) ,
\end{equation}
is called the \emph{symplectic matrix}. In this context, it can be
proved that any operator $A$ transforming $n$-mode states to
$n$-mode states can be expressed as
\begin{equation}
A=(2\pi)^{-n}\int \chi_A(\xi)W_{-\xi}d^{2n}\xi ,
\end{equation}
where $\chi_A(\xi)$ is the so called characteristic function and
$W_{\xi}$ are the Weyl operators, defined as
\begin{equation}
W_{\xi}=e^{i\xi^T\sigma R} ,\label{grupos}
\end{equation}
and $R=(R_1...R_{2n})$. Weyl operators satisfy the well-known
\emph{Weyl relations}
\begin{equation}
W_{\xi}W_{\eta}=e^{-i\xi^T\sigma\eta}W_{\xi+\eta} .
\end{equation}
When $A$ corresponds to the density operator associated to a
certain state, $\chi_A(\xi)$ is called the \emph{characteristic
function of the state} $A$. One can also define the \emph{Wigner
function} ${\cal W}_A(\xi)$ of $A$ as
\begin{equation}
{\cal W}_A(\xi)=(2\pi)^{-2n}\int
e^{i\xi·\sigma·\eta}\chi_A(\eta)d^{2n}\eta .
\end{equation}
The Wigner function behaves as a quasi probability distribution in
phase space. It is normalized, and integrating over $X_i$ or $P_i$
for each mode gives the corresponding probability distribution of
the remaining canonical variables.


For every state $\rho$, one can define its \emph{displacement
vector} $d$ as $d_k=tr\{\rho R_k\}$ and its \emph{covariance
matrix} $\gamma$ as $\gamma_{kl}=tr\{\rho
\{R_k-d_k,R_l-d_l\}_{+}\}$, where $\{\}_{+}$ denotes the
anti-commutator. Because of the Heisenberg uncertainty relations,
any state has to satisfy
\begin{equation}
\gamma\geq i\sigma \label{positividad} .
\end{equation}

\emph{Gaussian states} are those $n$-mode quantum states whose
characteristic function is of the form:
\begin{equation}
\chi(\xi)=e^{i\xi·\sigma·d-\xi·\sigma^T\gamma\sigma·\xi/4} .
\end{equation}
Thus, any Gaussian state is completely described by its
displacement vector $d$ and covariance matrix $\gamma$.


\emph{Gaussian operations} are completely positive maps
transforming Gaussian states into Gaussian states. Gaussian
operations were fully characterized in \cite{Cirac,Fiurasek}.
There, the authors show that a Gaussian state $G$ with covariance
matrix $\Gamma$ and displacement $\Delta$ can be associated to
each Gaussian operation ${\cal G}$. In particular, if $\Gamma$ and
$\Delta$ are given by
\begin{equation}
\Gamma=\left(\begin{array}{cc}\Gamma_1&\Gamma_{12}\\
\Gamma_{12}^T&\Gamma_2\end{array}\right) \quad
\Delta=\left(\begin{array}{c}\Delta_1\\\Delta_2\end{array}\right)
,
\end{equation}
then the application of ${\cal G}$ on a Gaussian state
$(\gamma,d)$ produces a Gaussian state $(\gamma',d')$ such that
\begin{eqnarray}
\label{gmap}
  \gamma' &=& \tilde{\Gamma}_1-\tilde{\Gamma}_{12}\frac{1}
  {\tilde{\Gamma}_2+\gamma}\tilde{\Gamma}_{12}^T \nonumber\\
  d' &=& \Delta_1+\tilde{\Gamma}_{12}\frac{1}
  {\tilde{\Gamma}_2+\gamma}(\Delta_2+d) ,
\end{eqnarray}
where $\tilde{\Gamma}=(\one\oplus \theta)\Gamma(\one\oplus
\theta)$ and $\theta=D(1,-1,1,-1...)$ is the transformation that
changes the sign of the momenta. Throughout this article,
$D(a,b,\ldots)$ will denote a diagonal matrix with non-zero
entries $a$, $b$ and so on.


A fundamental Gaussian operation is homodyne detection, that is,
the physical measurement of one of the canonical coordinates. Let
$\gamma$ define a Gaussian state with zero displacement vector.
Suppose $\gamma$ can be divided into modes as
\begin{equation}
\label{gammamod}
\gamma=\left(\begin{array}{cc}A&C\\C^T&B\end{array}\right) .
\end{equation}
If we measure the $X$ component of each of the modes corresponding
to $A$, obtaining the result $(X_1,X_2,...)$, system $B$ will turn
into a Gaussian state with covariance matrix \cite{Eisert}
\begin{equation}
B'=B-C^T(X A X)^{MP}C , \label{homodinac}
\end{equation}
and displacement vector
\begin{equation}
d_B=C^T(X A X)^{MP}d_A , \label{homodinad}
\end{equation}
where $d_A=(X_1,0,X_2,0...)$, $MP$ denotes the pseudo-inverse
(inverse on the range) and $X$ is the projector
$X=D(1,0,1,0,1,0...)$.


Another important subset of Gaussian operations is constituted by
the so-called \emph{symplectic transformations}. It can be proved
that unitary Gaussian operations are the ones that transform the
canonical coordinates in the following way:
\begin{equation}
R'=S·R+T ,
\end{equation}
where $T$ is a vector and $S$ is a matrix belonging to the
symplectic group $Sp\,(2n,{\cal R})$. The symplectic group is
given by those matrices leaving invariant the symplectic matrix,
i.e. satisfying $S\sigma S^T=\sigma$. When $T=0$, the
transformation is called symplectic. Under symplectic
transformations, the displacement vector and the covariance matrix
change into $d'=S·d$ and $\gamma'=S\gamma S^T$. Symplectic transformations are very relevant because
of the following


\textbf{Theorem (Williamson) \cite{Williamson}:} For any real and
positive definite $2n\times 2n$ matrix, $C$, one can find a
symplectic matrix $S$ such that
\begin{equation}
SCS^T=\oplus_{i=1}^n \lambda_i\one_2 ,
\end{equation}
where $\lambda_i>0$ are called the \emph{symplectic eigenvalues}
of $C$.


Because of (\ref{positividad}), if we apply this theorem to the
covariance matrix of a certain state, we will get that all its
symplectic eigenvalues $\lambda_i$ have to be greater or equal
than one. Moreover, for a Gaussian state with covariance matrix
$\gamma$, the identity $\tr(\rho^2)=\det(\gamma)^{-1/2}$ holds
(recall that $\tr(\rho^2)$ gives a measure of the purity of
$\rho$). So, a Gaussian state is pure if and only if all its
symplectic eigenvalues are equal to one.


Finally, let us give some known results about entanglement and
Gaussian states that will next be used. In this case, one
considers Gaussian states in bipartite systems of $n+m$ modes,
where Alice and Bob's systems are of $n$ and $m$ modes,
respectively.

\textbf{Theorem \cite{Werner}:} Let $\gamma_{AB}$ be the
covariance matrix of a Gaussian state in a bipartite system. This
state is separable if and only if
\begin{equation}
\gamma_{AB}\geq \gamma_a\oplus\gamma_b , \label{entangled}
\end{equation}
for certain \emph{physical} covariance matrices $\gamma_a$ and
$\gamma_b$ in systems $A$ and $B$, respectively.

Partial transposition is a positive, but not completely positive,
map that plays a key role in entanglement theory. In the case of
continuous variable systems, after partial transposition on, say,
system $B$, the sign of Bob's momenta is changed while the rest of
canonical coordinates is kept unchanged. At the level of
covariance matrices, this means that
$\gamma_{AB}\rightarrow\gamma'_{AB}=\theta_B\gamma_{AB}\theta_B$.
Therefore, a state $\rho$ has non-positive partial transposition
(NPPT) when $\gamma'_{AB}$ does not define a positive operator,
that is
\begin{equation}
\gamma_{AB}\not\geq i\tilde{\sigma} , \label{oso}
\end{equation}
where $\tilde{\sigma}=(\one_A\oplus \theta_B)\sigma(\one_A\oplus
\theta_B)$. It can be shown that this condition is equivalent to
\begin{equation}
\gamma_{AB}\not\geq \tilde{\sigma}\gamma_{AB}^{-1}\tilde{\sigma}^T
.\label{NPPT}
\end{equation}
The positivity of partial transposition, also known as PPT
criterion, represents a necessary and sufficient condition for
separability for $1\times 1$ \cite{11mode} and $1\times N$
Gaussian states \cite{Werner}, while it is only a necessary
condition for the rest of systems \cite{Werner}. It also gives a
necessary and sufficient condition for entanglement
distillability: a Gaussian state is distillable if and only if it
is NPPT \cite{GDCZ}.

\section{Secret bits from Gaussian states}
\label{cprivacy}

In our quantum cryptographic scenario, there are two parties,
Alice and Bob, who share several copies of a certain Gaussian
state, $\rho_{AB}$. As said, it is assumed that the honest parties
know to have $N$ independent copies of $\rho_{AB}$. There is also
an eavesdropper, Eve, that keeps the purification of that state.
In a prepare and measure scheme, the assumption in the state
preparation means that Eve interacts identically, individually and
in a Gaussian way with the states sent to Bob by Alice. Alice and
Bob perform some individual measurements over their copies and
afterwards apply Advantage Distillation, Error Correction and
Privacy Amplification techniques to extract a perfect secret key.
These three processes constitute the reconciliation part of the
protocol. We consider two type of attacks: (i) \emph{individual},
where Eve performs individual measurements, possibly non-Gaussian,
over her set of states before Alice and Bob's public
reconciliation, or (ii) \emph{collective}, where Eve waits until
the reconciliation is finished and then decides what (possibly
collective) measurement gives her more information on the final
key. Note that this second type of attacks is the most general
under the mentioned assumption in the state preparation. On the
other hand, to assume that Eve measures her state before the
reconciliation process, as for individual attacks, appears quite
reasonable from an experimental point of view.

In this section, we first review the protocol described in
\cite{mayo}. There, it was proved that (i) a secret key can be
distilled from any NPPT Gaussian state, provided that Eve is
restricted to individual attacks, (ii) there exist slightly
entangled states that become insecure, with our protocol, when
Eve's attack is collective and (iii) key distillation secure
against collective attacks is still possible for sufficiently
entangled states. Here, we will first improve the security
analysis against collective attacks, giving a necessary and
sufficient condition for secret key distillation from Gaussian
states with our protocol. Later, we will show that our scheme does
not allow to extract a secret key out of bound entangled Gaussian
states.

\subsection{Key distillation protocol}

The key distillation protocol presented in \cite{mayo} consists of
the following steps:

\begin{enumerate}
    \item Starting from $\rho_{AB}$, Alice and Bob apply the GLOCC
    protocol of \cite{GDCZ} mapping any NPPT Gaussian state of
    $n+m$ modes into an NPPT $1\times 1$ Gaussian and
    symmetric state, whose CM, see Eq. (\ref{gammamod}), is
    \begin{equation}
    \label{symmst}
    A=B=\begin{pmatrix}
    \lambda & 0 \cr
    0 & \lambda
    \end{pmatrix} \quad\quad
    C=\begin{pmatrix}
    c_x & 0 \cr
    0 & -c_p
    \end{pmatrix} ,
    \end{equation}
    where $\lambda\geq 0$ and $c_x\geq c_p\geq 0$. The positivity
    condition reads $\lambda^2-c_xc_p-1\geq \lambda(c_x-c_p)$ while
    the entanglement (NPPT) condition gives
    \begin{equation}
    \label{entcond}
    \lambda^2+c_xc_p-1< \lambda(c_x+c_p) .
    \end{equation}
    \item Each of them measure the $X$ quadratures of their modes,
    $X_A,X_B$. As soon as all measurements are done, Alice randomly chooses a real number $X_0>0$ and sends it to Bob via a classical channel. If their measured quadratures satisfy $|X_A|=|X_B|=X_0$, they accept the results. Otherwise, they
    discard them. They then make binary these results according to the
    prescription $X_i=X_0\rightarrow 0, X_i=-X_0\rightarrow 1, i=A,B$,
    thus obtaining a list of correlated bits.
    \item Alice and Bob apply Classical Advantage Distillation
    \cite{Maurer} over their lists of symbols: they randomly choose a set of $N$
    indices, and build binary $N$-vectors with the corresponding
    symbols appearing in their lists: $(A_1,A_2,...,A_N)$ for Alice and
    $(B_1,...,B_N)$ for Bob. Then, Alice generates a random bit,
    $c\in\{0,1\}$, and sends Bob a vector $(c_1,...,c_N)$ such that
    $A_1\oplus c_1=A_2\oplus c_2=...=A_N\oplus c_N=c$. Next, Bob
    computes the quantities $c_i'=B_i\oplus c_i$, $i=1,\ldots,N$. In case
    $c_1'=c_2'=...=c_N'=c'$, Bob accepts the symbol $c'$. Otherwise,
    he discards it. Anyhow, after this step Alice and Bob will have to
    throw away all the symbols used and repeat the process with the
    remaining symbols. At the end, they will have a reduced list of
    more correlated symbols.
    \item Alice and Bob apply Error Correction and Privacy Amplification
    protocols to the new list in order to obtain a secret key.
\end{enumerate}

Let us denote by $\omega_{AB}$ Alice and Bob's $1\times 1$ state
after step 1 and by $\epsilon_B$ the probability that Alice and
Bob obtain different results (namely, $(X_0, -X_0)$ or $(-X_0,
X_0)$) after the homodyne measurements and post-selection. Let us
also denote by $\ket{e_{\pm\pm}}$ Eve's resulting states when
Alice and Bob measure $(\pm X_0,\pm X_0)$. If Eve is restricted to
individual attacks, i.e. she measures before step 3, the honest
parties can distill a key when \cite{AMG}
\begin{equation}
\label{keycond}
    \frac{\epsilon_{B}}{1-\epsilon_{B}}<|\braket{e_{++}}
    {e_{--}}| .
\end{equation}
Actually, this security condition also holds for the case in which
Eve applies a measurement on a finite number of copies of her
states before the reconciliation process has started. As shown in
\cite{mayo}, Eq. (\ref{keycond}) is equivalent to demand that the
initial state $\rho_{AB}$ was NPPT.

Now, one would naturally wonder how this security condition has to
be modified when Eve is allowed to perform a collective attack,
i.e. she can measure after the public reconciliation. In this
case, Eve's information during the whole protocol is quantum. Note
that, once the honest parties accept a symbol after Advantage
Distillation, they can agree to both change its sign or not. This
is so because the symplectic transformation $(X_A,P_A, X_B,P_B)
\rightarrow (-X_A,-P_A,-X_B ,-P_B)$ leaves the Gaussian state
$\omega_{AB}$ invariant. Therefore, we can consider that Alice's
$N$ symbols employed in a successful performance of step 3 are
equal, and so Bob's. That is, the global state resulting from step
3 is (see also \cite{Bae})
\begin{eqnarray}\label{ccqcorr}
    \rho_{ABE}&=&\frac{1-\epsilon_{BN}}{2}\left([00]\otimes[e_{++}]^{\otimes
    N}
    +[11]\otimes[e_{--}]^{\otimes N}\right) \nonumber\\
    &+&\frac{\epsilon_{BN}}{2}\left([01]\otimes[e_{+-}]^{\otimes N}
    + [10]\otimes[e_{-+}]^{\otimes N}\right)
\end{eqnarray}
where $[\psi]$ denotes the projector onto $\ket{\psi}$ and
$\epsilon_{BN}$ is Bob's error probability after Advantage
Distillation. For large $N$, this error has the form \cite{AMG}
\begin{equation}\label{errbN}
    \epsilon_{BN}\propto\left(\frac{\epsilon_B}{1-\epsilon_B}\right)^N
    .
\end{equation}

In step 4, Alice and Bob apply the one-way key distillation
protocol given in \cite{Winter}. This protocol deals with the case
where Alice has a classical random variable, $A$, correlated to a
quantum state on Bob and Eve's hands, $\rho_{B|A}$ and
$\rho_{E|A}$. The achievable key rate satisfies \cite{Winter}
\begin{equation}
K_\rightarrow\geq \chi(A:B)-\chi(A:E)
\end{equation}
where $\chi(X:Y)$ denotes the Holevo bound \cite{holevo}, i.e.,
$\chi(X:Y)=S(\rho_Y)-\sum_X p\,(X)S\,(\rho_{Y|X})$ and
$\rho_Y=\sum_Xp\,(X)\rho_{Y|X}$. In our case, Alice and Bob have
classical variables, so $\chi(A:B)$ is actually equal to the
mutual information $I(A:B)$, which is a function of
$\epsilon_{BN}$. Let us compute in what follows $\chi(A:E)$.

Notice that in the limit of large $N$, the error terms in
(\ref{ccqcorr}) can be neglected, since $\epsilon_{BN}\rightarrow
0$. This means that the states $\rho_{E|A}$ are actually pure, so
$\chi(A:E)\approx S(\rho_E)$ for large $N$. If the covariance
matrix associated to the state $\omega_{AB}$ is given by
\begin{equation}
\gamma_{AB}=\left(\begin{array}{cc}\gamma_x&R\\R^T&S\end{array}\right)
,
\end{equation}
where $\gamma_x$, $R$, $T$ and $S$ are $2 \times 2$ matrices, one
can see that for large $N$, $S(\rho_E)\propto k_E^N$, where
\begin{equation}
\log k_E=-(X_0,X_0)(S-R^T\gamma_x^{-1}R-\gamma_x^{-1})
\left(\begin{array}{c}X_0\\X_0\end{array}\right).
\end{equation}
Actually, one has that $S-R^T\gamma_x^{-1}R-\gamma_x^{-1}=
\{(\sigma\gamma_{AB}^{-1}\sigma^T)_x\}^{-1}-\gamma_x^{-1}$.
Throughout this article, $(M)_x$ denotes the projection of a
generic operator $M$ onto the $x$ space. It follows from this
expression that $k_E=|\bra{e_{++}}e_{--}\rangle|^2$. Comparing now
the two quantities, it is clear that a positive key rate is
possible when
\begin{equation}
\label{collcond}
\frac{\epsilon_B}{1-\epsilon_B}<|\bra{e_{++}}e_{--}\rangle|^2
\end{equation}
This gives a sufficient condition for distilling a secret key. On
the other hand, if Eve applies the particular attack proposed in
\cite{Kaszlikowski}, our protocol turns out to be insecure if Eq.
(\ref{collcond}) does not hold \cite{mayo}. That is, Eq.
(\ref{collcond}) is indeed the necessary and sufficient condition
for positive key extraction using our GLOCC protocol from Gaussian
states. Therefore, this closes the security gap left open in the
analysis of \cite{mayo} (see also Fig. 1).

This result could somehow be expected: the application of the
projectors $[\pm X_0]$ transforms the original Gaussian state into
an effective 2-qubit state that tends to a Bell diagonal state in
the limit $N$ going to infinite. The necessary and sufficient for
positive key extraction from a two-qubit state has recently been
derived in \cite{Bae}. The bound given there looks identical to
(\ref{collcond}).

\begin{figure}
  \includegraphics[width=7cm]{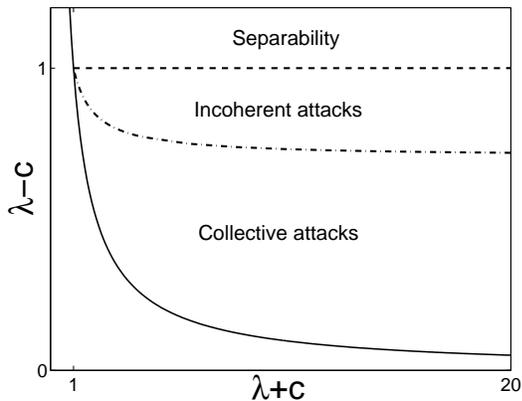}\\
  \caption{Security analysis of isometric
  $1\times 1$ Gaussian states, with covariance matrix satisfying,
  see Eq. (\ref{symmst}), $c_x=c_p=c$.
  All physical states  are above the solid line. The dashed line
  defines the entanglement limit, that coincides with the security
  bound against incoherent attacks. States below the dashed-dotted
  line are secure against any attack. It is implicitly assumed that
  Alice and Bob measure the $X$ quadratures.}
\end{figure}

\subsection{Bound entangled states}
\label{besubsec}

Our next result concerns the distillation of secret bits from PPT
Gaussian states using the previous GLOCC protocol. Recall that in
the Gaussian scenario, a state is entanglement distillable if and
only if it is NPPT. This means that there do not exist NPPT bound
entangled states. It is quite clear that the considered protocol,
in the form previously presented, does not allow to extract a
secret key from any PPT state. Indeed, in step 1 any PPT state is
mapped into a $1\times 1$ PPT state, which is separable
\cite{11mode}, and no secret key can be extracted from separable
states \cite{sep}. This is why we consider a generalized version
of the protocol above, where step 1 is replaced by: 1.' Alice and
Bob perform any GLOCC pre-processing, possibly non-deterministic,
over their states. Then, they measure the $X$ quadrature of one of
their modes as in step 2 and the protocol proceeds as explained
above. It is next shown that even in this more general scenario
and restricting Eve to an individual attack, no secret key
distillation is possible from PPT states.


As above, $\epsilon_B$ defines Bob's error probability after
homodyne measurement and postselection. Let $\rho_+$ and $\rho_-$
be Eve's resulting states when Alice and Bob measure $(X_0,X_0)$
or $(-X_0,-X_0)$, respectively. Contrary to the previous
situation, these states can now be mixed. Then, if Eve is
restricted to individual attacks, a secret key can be distilled
using our scheme if and only if
\begin{equation}
\frac{\epsilon_B}{1-\epsilon_B}<\tr\left(\sqrt{\sqrt{\rho_-}
\rho_+\sqrt{\rho_-}}\right) .\label{condition}
\end{equation}
It is possible to derive this formula from \cite{AMG}. There, it
is shown that Eve's error probability behaves as
$\epsilon_{EN}\propto
(\sum_{i=1}^{M}\sqrt{\tr(\rho_+M_i)\tr(\rho_-M_i)})^N$, where
$M_i$ is the $i^{th}$ operator corresponding to the $i^{th}$
outcome of Eve's measurement, $\sum_i M_i=\one$. Now, one has to
take into account that the minimum of $\sum_i
\sqrt{\tr(\rho_+M_i)\tr(\rho_-M_i)})$ over all possible
measurements is equal to the Uhlmann's fidelity \cite{uhlman} of
$\rho_+$ and $\rho_-$, namely
$\tr(\sqrt{\sqrt{\rho_-}\rho_+\sqrt{\rho_-}})$. A derivation of
this result can be found in \cite{Barnum}. Recall that Bob's error
probability after step 3 goes as (\ref{errbN}). Thus, for Alice
and Bob to extract a secret key it is enough that $\epsilon_{BN}$
decreases exponentially faster than $\epsilon_{EN}$. Then,
condition (\ref{condition}) immediately applies.


Our goal is now to express (\ref{condition}) in terms of
$\gamma_{AB}$. In fact, it will be seen that (\ref{condition}) is
equivalent to the NPPT condition for Gaussian states.


As usual, it is supposed that Eve's state is entangled with Alice
and Bob's one, so that the whole state is pure. Let's assume that
Alice and Bob have just finished the GLOCC pre-processing of step
1', and let's call $\gamma_{AB}^{(r)}$ the resulting reduced
covariance matrix that contains only their first modes. We
introduce the following notation:
\begin{eqnarray}
\label{descomp}
  \gamma_{ABE} &=& \begin{pmatrix}
  \gamma_{AB}^{(r)}& L& E \cr
  L^T& \gamma_{AB}^{(m)} & G\cr
  E^T& G^T& \gamma_E\end{pmatrix} \quad
  F=\begin{pmatrix}E\cr G\end{pmatrix} \nonumber\\
  \gamma_{AB}^{(r)} &=& \begin{pmatrix}\gamma_x & R\cr
  R^T& S\\\end{pmatrix} \quad
  (\gamma_{AB}^{(r)})^{-1}=\begin{pmatrix}X&Y\cr
  Y^T&Z\end{pmatrix} ,
\end{eqnarray}
%
%
%
%
%
where $\gamma_x$ and $X$ correspond to the $X_A,X_B$ space. The
following formula will next be useful \cite{Horn}:

\begin{eqnarray}
\label{matriz}
&\begin{pmatrix}A&C\cr C^T&B\end{pmatrix}^{-1}=& \nonumber\\
&\begin{pmatrix}(A-C\frac{1}{B}C^T)^{-1}& A^{-1}C(C^T
\frac{1}{A}C-B)^{-1})\cr(C^T\frac{1}{A}C-B)^{-1})C^TA^{-1}&
(B-C^T\frac{1}{A}C)^{-1}\end{pmatrix}. \nonumber\\
\end{eqnarray}
%
%
Using equations (\ref{homodinac}) and (\ref{homodinad}), it is
straightforward to check that $\rho_+$ is described by:
\begin{eqnarray}
\label{gamma}
\gamma_E'&=&\gamma_E-E^T\beta E  \nonumber\\
d_E'&=&E^T\beta
\begin{pmatrix}X_0 \\ X_0\\0\\0\end{pmatrix} ,
\end{eqnarray}
%
where
\begin{equation}
\beta=\left(\begin{array}{cc}\gamma_x^{-1}&0\\0&0\end{array}\right)
.
\end{equation}
Similarly, if Alice and Bob measure $-X_0$, Eve's corresponding
state $\rho_-$ will have the same covariance matrix and opposite
displacement vector.


Let us first calculate the right hand side of (\ref{condition}).
It can be shown (see Appendix A) that
\begin{equation}
\tr(\sqrt{\sqrt{\rho_-}\rho_+\sqrt{\rho_-}})=e^{-d_E^{'T}\gamma_E^{'-1}d_E'}
.\label{apendicitis}
\end{equation}
Now we want to write this in terms of $\gamma_{AB}$. If we define
\begin{equation}
    E=\left(\begin{array}{c}E_x\\E_p\\\end{array}\right) ,
\end{equation}
where $E_x$ is the part of $E$ corresponding to the $X$
quadratures, we only have to substitute to get that
$d_E^{'T}\gamma_E^{'-1}d_E'$ can be written as
\begin{equation}
\left(\begin{array}{cc}X_0&X_0\\\end{array}\right)
  \gamma_x^{-1}E_x(\gamma_E-E_x^T\gamma_x^{-1}E_x)^{-1}E_x^T
  \gamma_x^{-1}\left(\begin{array}{c}X_0\\X_0\\\end{array}\right)
  .
\end{equation}
Using formula (\ref{matriz}) applied to the matrix
\begin{equation}
    K=\left(\begin{array}{cc}\gamma_x&E_x\\E_x^T&\gamma_E\\\end{array}\right)^{-1}
    ,
\end{equation}
and the condition $KK^{-1}=\one$, we have that
$E_x(\gamma_E-E_x\gamma_x^{-1}E_x)^{-1}E_x^T
\gamma_x^{-1}=\gamma_x(\gamma_x-E_x\gamma_E^{-1}E_x^T)^{-1}-\one$.
Substituting, we arrive at
\begin{eqnarray}
  &d_E^{'T}\gamma_E^{'-1}d_E'=  \nonumber\\
  & \left(\begin{array}{cc}X_0&X_0\\\end{array}\right)
  ((\gamma_x-E_x\gamma_E^{-1}E_x^T)^{-1}-\gamma_x^{-1})
  \left(\begin{array}{c}X_0\\X_0\\\end{array}\right) .
\end{eqnarray}
Note that $(\gamma_x-E_x\gamma_E^{-1}E_x^T)$ is just the
projection of $(\gamma_{AB}-F\gamma_E^{-1}F^T)$ onto the $x$
space. Therefore, one can replace in the previous expression
$(\gamma_x-E_x\gamma_E^{-1}E_x^T)$ and $\gamma_x^{-1}$ by
$(\gamma_{AB}-F\gamma_E^{-1}F^T)_x$ and $(\gamma_{AB}^{-1})_x$.

On the other hand, we have assumed that Eve purifies the state
shared by Alice and Bob. Since all purifications are equivalent up
to a unitary transformation on Eve's space, one can consider a
particular purification without loosing generality. One possible
purification, see (\ref{descomp}), is given by \cite{Giedke}
\begin{equation}
    F=\sigma_{AB}[-(\sigma_{AB}\gamma_{AB})^2-\one]^{1/2}
  \theta \quad \gamma_E = \theta\gamma_{AB}\theta .
\end{equation}
If $S$ is the symplectic matrix such that $S^T\gamma_{AB}S$ is
diagonal, one can verify that
\begin{eqnarray}
  F\gamma_E^{-1}F^T &=& -\sigma_{AB}S(\oplus_k
  \sqrt{\lambda_k^2-1}\one_2)S^{-1}S
  (\oplus_k\frac{1}{\lambda_k}\one_2) \nonumber\\
  &&S^T(S^{-1})^T(\oplus_k\sqrt{\lambda_k^2-1}\one_2)
  S^T\sigma_{AB} \nonumber\\
  &=& -\sigma_{AB}S(\oplus_k\frac{\lambda^2_k-1}
  {\lambda_k}\one_2)S^T\sigma_{AB} \nonumber\\
  &=&\gamma_{AB}-\sigma_{AB}\gamma_{AB}^{-1}\sigma_{AB}^T .
\end{eqnarray}
So $\gamma_{AB}-F\gamma_EF^T=\sigma\gamma_{AB}^{-1}\sigma^T$ and
$\tr(\sqrt{\sqrt{\rho_-}\rho_+\sqrt{\rho_-}})$ is equal to
\begin{equation}
\exp\left[-X_0^2\left(\begin{array}{cc}1&1\\\end{array}\right)
\left((\sigma_{AB}\gamma_{AB}^{-1}\sigma_{AB}^T)_x^{-1}-
(\gamma_{AB})_x^{-1}\right)
\left(\begin{array}{c}1\\1\\\end{array}\right)\right] .
\label{whitelabel}
\end{equation}
The next step is to calculate the left hand side of
(\ref{condition}).

Let $\rho(X_A,X_B)$ be the probability density of $(X_A,X_B)$, the
$X$ quadratures of the reduced state $\gamma^{(r)}$. The
corresponding Wigner function satisfies:
\begin{equation}
    w(\xi)\propto e^{-\xi^T(\gamma_{AB}^{(r)})^{-1}\xi} .
\end{equation}
If $\xi=(x^A_1,x^B_1,\vec{p})$, then, according to
(\ref{descomp}),
\begin{eqnarray}
  \rho(x^A_1,x^B_1) &\propto& \int e^{-(\xi^T(
  \gamma_{AB}^{(r)})^{-1}\xi)}d\vec{p} \nonumber\\
  &=& \int
  e^{-(\vec{x}^TX\vec{x}+2\vec{x}^TY\vec{p}+\vec{p}^TZ\vec{p})}d\vec{p}
  .
\end{eqnarray}
Finally, we get
\begin{equation}
    \rho(x^A_1,x^B_1)\propto e^{-\vec{x}^T(X-YZ^{-1}Y^T)\vec{x}} .
\end{equation}
But, looking at (\ref{matriz}), we see that this is just
$\exp({-\vec{x}^T\gamma_x^{-1}\vec{x}})$. Writing
\begin{equation}
\gamma_x=\left(\begin{array}{cc}a&b\\b&c\\\end{array}\right) ,
\end{equation}
it is easy to see that, in our protocol,
\begin{equation}
    \frac{\epsilon_B}{1-\epsilon_B}=\exp\left(-\frac{4bX_0^2}
    {ac-b^2}\right) .
\end{equation}
%
%
In a similar way, one can define
\begin{equation}
(\sigma\gamma_{AB}^{-1}\sigma^T)_x=\left(\begin{array}{cc}d&e\\e&f\\\end{array}\right)
,
\end{equation}
and then, the term in the exponent of Eq. (\ref{whitelabel}) can
be expressed as
\begin{equation}
    -X_0^2(\frac{d-2e+f}{de-f^2}-\frac{a-2b+c}{ac-b^2}) .
\end{equation}

Collecting all these results, the condition (\ref{condition}) for
distilling a key with this protocol is equivalent to
\begin{equation}
\frac{d+f-2e}{df-e^2}-\frac{a+c+2b}{ac-b^2}<0 \label{condi} .
\end{equation}
We are now in a position to prove the next

\textbf{Theorem:} A secret key secure against individual attacks
can be distilled with our GLOCC protocol from a Gaussian state if
and only if the state is NPPT.

{\sl Proof:} The idea of the proof is to show that condition
(\ref{condition}) is equivalent to the PPT criterion. 
First, note that Eq. (\ref{condi}) can be rewritten as
\begin{equation}
    \left(\begin{array}{cc}1&-1\\\end{array}\right)(
    (\tilde{\sigma}\gamma_{AB}^{-1}\tilde{\sigma}^T)_x^{-1}-\gamma_x^{-1})
    \left(\begin{array}{c}1\\-1\\\end{array}\right) < 0
\end{equation}
Since $(\tilde{\sigma}\gamma_{AB}^{-1}\tilde{\sigma}^T)_x$ and
$\gamma_x$ are positive operators, the previous equation implies
that
$\gamma_{AB}-\tilde{\sigma}\gamma_{AB}^{-1}\tilde{\sigma}^T\not\geq
0$. But this is the condition for a Gaussian state to be NPPT, as
stated in (\ref{NPPT}). Therefore, if a key can be distilled out
of a Gaussian state with the previous protocol, this state has to
be NPPT. For the opposite implication one simply has to apply the
protocol of \cite{mayo}, that has previously been described.




\section{CONCLUSIONS}
\label{concl}


In this article, we have analyzed the extraction of secret bits
from quantum states in the every-day-growing field of Quantum
Information Theory with Continuous Variables. We have first
reviewed the protocol and results of \cite{mayo}: a secret key can
be distilled from any NPPT state when Eve is restricted to
individual attacks. In the more general scenario of collective
attacks, we extend the analysis of \cite{mayo}, providing a
necessary and sufficient condition for key distillability, with
the considered protocol. This protocol turns out to be completely
useless for bound entangled states, even in the case of individual
attacks. Before concluding, we would like to discuss several open
questions and implications that follow from our results.


First, note that all the presented results aim at answering
whether secret bits can be extracted from Gaussian states by
GLOCC. In terms of resources, we study the conversion of Gaussian
states into secret bits. However, very little is said about the
rate governing this conversion. This problem appears as a natural
follow-up of the present work. Notice that, strictly speaking, the
considered key-distillation protocol has zero rate. Indeed, the
probability that Alice and Bob obtain the outcomes $\pm X_0$ is
zero. Of course, the analysis can easily be adapted to a protocol
with finite rate: Alice and Bob only have to accept outcomes in
the range $|X_0\pm\delta|$, where $\delta>0$. By choosing a
properly small $\delta$, the security conditions still hold
because of continuity, while the protocol automatically acquires a
finite rate. It is intriguing the fact that both security
conditions, Eqs. (\ref{keycond}) and (\ref{collcond}), are
independent of $X_0$. This suggests that key distillation should
still be possible when Alice and Bob directly assign a bit to the
sign of their measurements, without discarding any value. This
would represent a significant improvement of the final key-rate.
Unfortunately, this result remains unproven. It would also be
desirable to adapt the reconciliation process to the continuous
character of the measured quantity, in a similar way as the
sliced-reconciliation protocols for error correction introduced in
\cite{van Assche}.

Another related question is the distribution of quantum states.
All our results were based on the hypothesis that Alice and Bob
share $N$ independent realization of the same known Gaussian
state. However, in any practical cryptographic protocol, Alice and
Bob will send and measure quantum states through an insecure
channel. From the observed probabilities, they have to infer what
their correlations with the environment are. This is indeed a very
delicate process that has not been considered here. For instance,
the honest parties cannot in principle exclude the existence of
correlations between the different quantum systems they measure.
While in our case, we simply assume that $N$ copies of the same
Gaussian state were given as an initial resource.

At a more fundamental level, our analysis represents one of the
first steps in the identification of the set of Gaussian states
that can be converted into secret bits by GLOCC. As discussed in
\cite{mayo}, for any Gaussian state one can define $GK_D$ and
$GE_D$, quantities that specify the amount of secret and entangled
bits extractable from it by GLOCC protocols. The results of Refs.
\cite{Cirac,Eisert,Fiurasek} imply that $GE_D=0$. On the other
hand, it follows from \cite{mayo} and this work that $GK_D$ is
non-zero for sufficiently entangled NPPT states. It would be
relevant to extend the present results, proving that $GK_D>0$ for
some states violating our security conditions. An almost
unexplored possibility in this direction is the use of global, but
still Gaussian, operations by Alice and Bob. In particular, note
that in the analyzed protocol, all the quantum operations were at
the single-copy level. Therefore, it is unknown whether the use of
coherent quantum operations gives any improvement for key
extraction. A related open question is the existence of the
so-called ``entanglement purification" protocols \cite{Cirac},
where Alice and Bob map many copies of a noisy entangled state
into a pure entangled state (not necessarily maximally entangled).
The goal would then be to decouple the honest parties' correlation
from the eavesdropper, something that it is sufficient in a
cryptographic scenario.

The case of bound entangled states is also of particular interest. Indeed, our result suggest that $GK_D=0$ for
all these states (c.f. \cite{Horo}). In the same spirit as in Ref. \cite{Horo}, one could look for
\emph{Gaussian secret states}. These would be states for which there exist Gaussian measurements by Alice and
Bob almost perfectly correlated about which Eve has arbitrarily small information. The results of section
\ref{besubsec} rule out this possibility for PPT states. Indeed, if this were the case, there would be PPT
secret states. This would imply that our protocol would work for a PPT state, which has been shown here to be
impossible. Unfortunately, this does not allow to conclude that
$GK_D=0$ for PPT states. 
More in general, it would also be interesting to prove that
$K_D>0$ for a Gaussian PPT state, i.e., that key extraction is
possible, even if the distillation protocol employs a non-Gaussian
operation.


\section{Acknowledgements}

We acknowledge discussion with Ignacio Cirac, Jens Eisert and
G\'eza Giedke. This work has been supported by the Ministerio de
Ciencia y Tecnolog\'\i a, under the ``Ram\'on y Cajal" grant, and
the Generalitat de Catalunya.

\section*{APPENDIX: PROOF OF RELATION (\ref{apendicitis})}


From the definition of the characteristic function and relation
(\ref{grupos}) it can be derived that
\begin{equation}
    \rho_1^2\rightarrow \chi_1^{(2)}(\xi)=e^{-i\xi^T\sigma d_E'
     -\xi^T\sigma D(\gamma_E')\sigma^T\xi/4}S(\gamma_E') .
\end{equation}
Therefore,
\begin{equation}
    \sqrt{\rho_1}\rightarrow \chi_1^{(1/2)}(\xi)=
    e^{-i\xi^T\sigma d_E' -\xi^T\sigma V(\gamma_E')
    \sigma^T\xi/4}H(\gamma_E') ,
\end{equation}
and then
\begin{equation}
    \sqrt{\rho_1}\rho_0\sqrt{\rho_1}\rightarrow
    A(\gamma_E',d_E')e^{-\xi^TB(\gamma_E')\xi +i\xi^T
    \sigma r(\gamma_E',d)} .
\end{equation}
Using the cyclic property of the trace, we get
$\tr(\sqrt{\rho_1}\rho_0\sqrt{\rho_1})=\tr(\rho_1\rho_0)=
e^{-2d_E^{'T}\gamma_E^{'-1}d_E'}|\gamma_E'|^{-1/2}=A(\gamma_E',d_E')$.
One then has
\begin{eqnarray}
    &\sqrt{\sqrt{\rho_1}\rho_0\sqrt{\rho_1}}\rightarrow \nonumber\\
    &C(\gamma_E')\sqrt{A(\gamma_E',d_E')}e^{-\xi^TU(\gamma_E')\xi
    +i\xi^T\sigma s(\gamma_E',d_E')} ,
\end{eqnarray}
which, after substitution, gives
\begin{equation}
    \tr(\sqrt{\sqrt{\rho_1}\rho_0\sqrt{\rho_1}})=
    e^{-d_E'^{'T}\gamma_E^{'-1}d_E'}M(\gamma_E')
\end{equation}
However, note that if $d_E'=0$, $\rho_0=\rho_1$, and
$\tr(\sqrt{\sqrt{\rho_0}\rho_0\sqrt{\rho_0}})=1$. This implies
that $M(\gamma_E')=1$, and so:
\begin{equation}
\tr(\sqrt{\sqrt{\rho_1}\rho_0\sqrt{\rho_1}})=
e^{-d_E^{'T}\gamma_E^{'-1}d_E'}
\end{equation}



\end{document}